\documentclass[10pt,twocolumn,twoside]{IEEEtran}
\usepackage{empheq,amssymb}
\usepackage{braket}
\usepackage{dsfont,mathrsfs}
\usepackage{enumerate}
\usepackage{graphicx}
\usepackage{amsmath}
\usepackage{amsthm}
\usepackage{stfloats}
\usepackage{color}

\newtheorem{definition} {Definition}

\newtheorem{theorem}    {Theorem}
\newtheorem{lemma}      {Lemma}
\newtheorem{corollary} {Corollary}
\newcommand\EXP[1]{\mathop{\kern0pt \mathds E}{\Set{#1}}}
\newcommand\PR [1]{\mathop{\kern0pt \Pr}{\Set{#1}}}

\newcommand{\leftexp}[2]%
  {\mathop{}%
   \mathopen{\vphantom{#2}}^{#1}%
   \kern-\scriptspace%
   #2}

\makeatletter
\newcommand\SEQ{\@ifstar\SEQB\SEQA}
\makeatother

\newcommand\SEQA[2][T]{\{#2_t$, $t=1,\dots,#1\}}
\newcommand\SEQB[1]{\{#1_1$, $t=1,\dots\}}

\title {\LARGE \bf Signaling in sensor networks for sequential detection }
\author{
Ashutosh Nayyar and Demosthenis Teneketzis 
\thanks{A.\ Nayyar is with Ming Hsieh Department of Electrical Engineering at University of Southern California, Los Angeles ({\tt\small ashutosn@usc.edu}).  D.\ Teneketzis is with Department of Electrical Engineering and Computer Science,
        University of Michigan, Ann arbor ({\tt\small teneket@eecs.umich.edu}). The second author's work was supported by NSF under grant number 11-11061 and by NASA under grant number NNX12AO54G.
         }
         }
\begin{document}
\maketitle

\begin{abstract}

Sequential detection problems in sensor networks are considered. The true state of nature/true hypothesis is modeled as a binary random variable $H$ with known prior distribution. There are $N$ sensors making noisy observations about the hypothesis; $\mathcal{N} =\{1,2,\ldots,N\}$ denotes the set of sensors. Sensor $i$ can receive messages from a subset $\mathcal{P}^i \subset \mathcal{N}$ of sensors and send a message to a subset $\mathcal{C}^i \subset \mathcal{N}$. Each sensor is faced with a stopping problem. At each time $t$, based on the observations it has taken so far and the messages it may have received,  sensor $i$ can decide to stop and communicate a binary decision  to the sensors in $\mathcal{C}^i$, or it can continue taking observations and receiving messages.  After sensor $i$'s binary decision has been sent, it becomes inactive. Sensors incur operational costs (cost of taking observations, communication costs etc.) while they are active. In addition, the system incurs a terminal cost that depends on the true hypothesis $H$, the sensors' binary decisions and their stopping times. The objective is to determine decision strategies for all sensors to minimize the total expected cost. Even though sensors only communicate their final decisions, there is implicit communication every time a sensor decides not to stop. This implicit  communication through decisions is referred to as signaling. The general communication structure results in complex signaling opportunities in our problem. In spite of the generality of our model and the complexity of signaling involved, it is shown that the a sensor's posterior belief on the hypothesis (conditioned on its observations and received messages) and its received messages constitute a sufficient statistic for decision making and that all  signaling possibilities are effectively captured by a  4-threshold decision rule where the thresholds depend on received messages. 

\end{abstract}
\section{Introduction}
The problem of decentralized detection with a group of sensors has received considerable attention in the literature. The basic problem structure involves (i) multiple sensors that can make observations about the environment, (ii) limited communication resources which prevent sensors from sharing all their information and (iii) the requirement to make a decision about a binary hypothesis about the environment. Typically, one of the sensors (or a non sensing entity) serves as a fusion center that receives messages from other sensors and makes a decision about the hypothesis. 

Within this basic structure, two classes of problems can be distinguished. In \emph{static} problems, each sensor makes  a fixed number of observations about the environment. It may also receive messages from other sensors. The sensor then quantizes this information and transmits a message to a fusion center and/or to other sensors. Such problems have been extensively studied since their initial formulation in \cite{Tenney_Detection} (See the surveys in \cite{Tsitsiklis_survey}, \cite{Varshney}, \cite{Tay_tsitsiklis}, \cite{Papastavrou} and references therein). Static problems with large number of sensors were studied in \cite{Tsitsiklis_large}, \cite{Chamberland04}, \cite{Papastavrou_2}. The key feature to note here is that the number of observations made by each sensor is fixed a priori. The second class of decentralized detection problems are \emph{sequential problems} where the number of observations made by each sensor is a stopping time. In one formulation of such problems, the fusion center makes the stopping decision for each sensor based on the information it gathers \cite{VBP_Detection}. In this formulation, even though the communication decisions are made in a decentralized manner (that is, each sensor decides what message should be sent to the fusion center), the stopping decisions are made in a centralized manner.  This formulation requires sensors to continuously send messages to the fusion center which decides when each sensor should stop.

The problem studied in this paper is motivated by situation when sensors cannot communicate continuously. Instead, each sensor makes its own stopping decision and communicates only once after making the decision to stop. Note that both the stopping and communication decisions are now decentralized. The decentralization of stopping decisions introduces new \emph{signaling} aspects in these problems that are absent from static problems and sequential problems where stopping decisions are centralized. Consider a sensor that has to first make a stopping decision and, after stopping, send a message to a fusion center.  At each time before the stopping time, the sensor's decision not to stop is implicitly observed by the fusion center. This decision conveys information about the observations that the sensor has made so far. This implicit communication through decisions is referred to as {signaling} in decision problems with multiple decision makers \cite{Ho_signaling, Spence}.

Sequential problems where sensors make their own stopping decisions but there is no possibility of  signaling were considered in \cite{Wald} (with only $1$ sensor) and in  \cite{Dec_Wald, LaVigna_86} (with multiple non-communicating sensors).
In all these settings, it was shown that two-threshold based stopping rules are optimal. Sequential problems where signaling was present were studied in \cite{Nayyar_MTNS} and \cite{NayyarTeneketzis:2009} respectively. It was shown that optimal decision strategies for sensors are characterized by $2M$ thresholds where $M$ is the size of the communication alphabet.
\subsection{Contributions}     
We consider a sequential decentralized detection problem where sensors make their own  decisions about how many observations to take and what message to send after stopping. We consider an arbitrary communication topology where sensor $i$ can send a message to an arbitrary (but fixed) subset $\mathcal{C}^i$ of sensors and receive messages from another subset $\mathcal{P}^i$ of sensors. The general communication topology distinguishes our work from similar problems with two sensors or with a star topology for communication \cite{Nayyar_MTNS}, \cite{NayyarTeneketzis:2009}. The communication structure results in complex signaling opportunities in our problem. For example,  sensor $j$ may be receiving message from sensor $1$ which in turn is receiving message from sensors in the set $\mathcal{P}^1$, which in turn may be receiving messages from other sensors. Thus, sensor $1$'s decision signals information about its own observations as well as the information it gathers  due to signaling by sensors in the set  $\mathcal{P}^1$. In spite of the generality of our model and the complexity of signaling involved, we will show that the a sensor's posterior belief on the hypothesis (conditioned on its observations and received messages) and its received messages constitute a sufficient statistic for decision making and that all signaling possibilities are effectively captured by a  4-threshold decision rule where the thresholds depend on received messages. 
\subsection{Organization}
         The rest of the paper is organized as follows. We formulate our problem in Section~\ref{sec:PF}. We present the information states (sufficient statistics) for the sensors in Section~\ref{sec:IS}. We give a counterexample that shows that classical two-thresholds are not necessarily optimal  in Section~\ref{sec:counter}. We derive a parametric characterization of optimal strategies in Section~\ref{sec:thresholds}. We conclude in Section~\ref{sec:con}.
\subsection{Notation}
Subscripts are used as time index and the superscripts are used as the index of the sensor.  \(X_{1:t}\) refers to the sequence \(X_1, X_2,..,X_t\).  For a collection of sensor indices $\mathcal{P} = \{a,b,\ldots,z\}$, the notation $X^{\mathcal{P}}_t$ refers to the collection of variables $X^a_t,X^b_t,\ldots,X^z_t$. We use capital letters to denote random variable and the corresponding lower case letters for their realizations. $\mathds{P}(\cdot)$ denotes probability of an event and $\mathds{E}[\cdot]$ denotes expectation of a random variable.  For a random variable $X$ and a realization $x$, we sometimes use $\mathds{P}(x)$ to denote the probability of event $\{X=x\}$. 

\section{Problem formulation} \label{sec:PF}
   We consider a binary hypothesis testing problem where the true hypothesis is modeled as a random variable \(H\) taking values 0 or 1 with known prior probabilities:
   \[\mathds{P}(H=0) = p_0; \hspace{10pt} \mathds{P}(H=1) = 1-p_0\]
 There are $N$ sensors, indexed by $i=1,2,\ldots,N$. Let $\mathcal{N} := \{1,2,\ldots,N\}$.  Each sensor can make noisy observations of the true hypothesis. Conditioned on the hypothesis \(H\), the following statements are assumed to be true: \\
1. The observation of the \(i^{th}\) sensor at time \(t\), \(Y^i_t\) (taking values in the set \(\mathcal{Y}^i\)), either has a discrete distribution  or admits a probability density function. For convenience, we will denote both the conditional discrete distribution and the conditional density by  $f^i_t(\cdot|H)$. \\
2. Observations of the \(i^{th}\) sensor at different time instants are conditionally independent given \(H\).\\
3. The observation sequences at different sensors are conditionally independent given \(H\). \\
%
%
\subsection{Communication and Decision Model}
Each sensor is faced with a stopping problem. At any time $t$,  sensor $i$, $i \in \mathcal{N}$, can decide to stop and make a binary decision --- $U^i_t= 0$ or $U^i_t=1$ --- or it can decide to continue operating which we will denote by $U^i_t=b~(blank)$. 

We assume that the communication model among the sensors is described by a directed graph $\mathcal{G} = (\mathcal{N}, \mathcal{E})$. A directed edge from sensor $i$ to sensor $j$ means that  the \emph{final decision} made at sensor $i$ is communicated to sensor $j$. Note that the observations of sensor $i$ are \emph{not} communicated to sensor $j$. This limited communication may be justified by resource constraints such as limited battery life for the sensors that prohibit continuous communication among them.

 We denote by $\mathcal{C}^i$ the set of all sensors that have access to sensor $i$'s decisions, that is,
\begin{equation}
 \mathcal{C}^i := \{j \in \mathcal{N} | (i,j) \in \mathcal{E}\}.
\end{equation}
Similarly, we define $\mathcal{P}^i$ to be the set of all sensors whose decisions sensor $i$ has access to, that is,
\begin{equation}
 \mathcal{P}^i := \{j \in \mathcal{N} | (j,i) \in \mathcal{E}\}.\label{eq:parents}
\end{equation}
If sensor $i$ does not receive any message from sensor $j \in \mathcal{P}^i$ at time $t$, it implies that $U^j_{t}= b$. 

At any time $t$,  the information available to sensor $i$ before it makes its decision is 
\begin{equation}
 I^i_t = \{Y^i_{1:t}, \{U^j_{1:t-1}\}_{j \in \mathcal{P}^i} \} =: \{Y^i_{1:t}, U^{ \mathcal{P}^i}_{1:t-1} \}.
\end{equation} 
Sensor $i$ can use this information to decide whether to stop taking measurements and decide $U^i_t =0$ or $U^i_t=1$ or to continue taking measurements, that is, $U^i_t=b$. Once the sensor makes a stopping decision (that is, $U^i_t =0$ or $1$), it becomes inactive and all further decisions are assumed to be $b$.


  If sensor $i$ has not already stopped before, its decision at time $t$ is chosen according to a decision rule $\gamma^i_t$,
 \begin{equation}
 U^i_t = \gamma^i_t(Y^i_{1:t},U^{\mathcal{P}^i}_{1:t-1}).
 \end{equation}
 The collection of functions $\Gamma^i:=(\gamma^i_t$, $t=1,2,\ldots)$ constitute the \emph{decision strategy} of the $i^{th}$ sensor.  We define the following stopping times:
\[\tau^i := \min\{t:U^i_t \neq b\}, ~~ 1 \leq i \leq N\]

We assume that each sensor must make its final decision  no later than a finite horizon $T$, hence we have that $\tau^i \leq T$  for all $i$.  At any time $t$,  we define the set of active sensors as:
\begin{equation} \label{eq:active}
 \mathcal{A}_t := \{j: \tau^j \geq t\}.
\end{equation} 

\subsection{System Cost}
Sensors incur operational costs while they are active. These include the cost of taking measurements and the cost of communicating with other sensors. The operational costs are a function of the stopping times of the sensors and are given by the function $O(\tau^1,\ldots,\tau^N)$. Further, the system incurs a terminal cost that depends on the hypothesis $H$, the final decisions made by the sensors $U^i_{\tau^i}$ and the stopping times $\tau^i$ of the sensors. We denote this terminal cost by $A(H,U^1_{\tau^1}, U^2_{\tau^2},\ldots,U^N_{\tau^N}, \tau^1,\tau^2,\ldots,\tau^N)$. 
Thus, the total cost to the system can be written as:
\begin{align*}
J(H, \{U^i_{\tau^i}, \tau^i\}_{i \in \mathcal{N}}) = &O(\tau^1,\ldots,\tau^N) \\ + &A(H, U^1_{\tau^1}, \ldots,U^N_{\tau^N}, \tau^1,\ldots,\tau^N).
\end{align*}
The system objective is to choose the decision strategies $\Gamma^1, \Gamma^2,\ldots, \Gamma^N$ to minimize the expected value of the system cost,
 \begin{align}
  &\mathcal{J}(\Gamma^1, \Gamma^2,\ldots, \Gamma^N) 
  := \mathds{E}^{\Gamma^1,\Gamma^2,\ldots, \Gamma^N}\Big[J(H, \{U^i_{\tau^i}, \tau^i\}_{i \in \mathcal{N}})\Big] \label{eq:objective}
 \end{align}    
  where the superscript $\Gamma^1,\Gamma^2, \ldots, \Gamma^N$ over the expectation denotes that the expectation is with respect to a measure that depends on the choice of the strategies $\Gamma^1,\Gamma^2, \ldots, \Gamma^N$.
  
 \subsection{Special Cases}
 By making suitable choices of the communication graph and the system cost function, our model can be reduced to several known models of decentralized detection problems.
 
 \subsubsection{No communication model} Consider the case where no sensor sends its final decision to any other sensor, that is, the communication graph is $\mathcal{G} = (\mathcal{N}, \emptyset)$. Further, the operational cost is linear in the sensors' stopping times and the terminal cost depends only on the terminal decisions and the true hypothesis. That is,
 \begin{equation}
J(H, \{U^i_{\tau^i}, \tau^i\}_{i \in \mathcal{N}}) = \sum_{i=1}^N c^i \tau^i+ A(H, U^1_{\tau^1}, U^2_{\tau^2},\ldots,U^N_{\tau^N}).
\end{equation}
 Such a model was considered in \cite{Dec_Wald} with $N=2$ sensors.
 
 \subsubsection{One-way communication} Consider the case where sensors $2$ to $N$  send there final decision to sensor $1$ which is responsible for making a final decision on the hypothesis. The system costs are given as
  \begin{equation}
J(H, \{U^i_{\tau^i}, \tau^i\}_{i \in \mathcal{N}}) = \sum_{i=1}^N c^i \tau^i+ A(H, U^1_{\tau^1}).
\end{equation}
Such a model was considered in \cite{NayyarTeneketzis:2009}. An extension of this model is the case where the communication graph is a tree with sensor $1$ as the root. A static detection problem with such a network was considered in \cite{Tay_tsitsiklis}.
 
 \subsubsection{Two-way communication} Consider the case of two sensors that can both communicate their final decision to the other. The decision of the sensor that stops at a later time is considered to be the final decision on the hypothesis. The system costs are given as
  \begin{equation}
J(H, U^1_{\tau^1}, \tau^1, U^1_{\tau^2}, \tau^2) = \sum_{i=1}^2  c^i\tau^i+ A(H, U^1_{\tau^1}, U^2_{\tau^2},\tau^1, \tau^2),
\end{equation}
where $A(H, U^1_{\tau^1}, U^2_{\tau^2},\tau^1, \tau^2)$ is given as
\begin{align}
A(H,U^1_{\tau^1}, U^2_{\tau^2},\tau^1,\tau^2) = \left\{ \begin{array}{ll}
                                                       a(H, U^1_{\tau^1}) & \mbox{if $\tau^1 \geq \tau^2$} \\
                                                       a(H, U^2_{\tau^2}) & \mbox{if $\tau^2 > \tau^1$}
                                                       \end{array}
                                                    \right.
\end{align}
where $a(x,y) := \mu\mathds{1}_{\{x \neq y\}}$ and $\mu >0$. Such a model was considered in \cite{Nayyar_MTNS}.
 
\subsection{Signaling}
Consider a simple two sensor network where sensor $1$ can communicate its binary decision to sensor $2$. In the static version of this problem, sensor $1$ makes an observation, quantizes it and sends it to sensor $2$. The message sent is a compressed summary of sensor $1$'s observation. In the sequential version of this problem, sensor $1$ can make multiple observations and it has to decide when to stop making further observations. When the sensor stops, it sends a binary message to sensor $2$. As in the static case, the message sent at the stopping time is a compressed version of the observations of sensor $1$. However, unlike the static problem, this final message is not the only means by which sensor $1$ conveys information to sensor $2$. At each time before the stopping time, sensor $1$'s decision not to stop is observed by sensor $2$ (since it does not receive the binary message from sensor $1$ at that time). This decision conveys information about the observations that sensor $1$ has made so far. This implicit communication through decisions is called \emph{signaling} in decision problems with multiple decision makers \cite{Spence, Ho_signaling}. It is the presence of signaling in sequential problems in decentralized detection of the kind formulated in this paper that distinguishes them from static problems.

The signaling is more complicated in the general problem formulated in Section \ref{sec:PF}. Sensor $2$ may be receiving message from sensor $1$ which in turn is receiving message from sensors in the set $\mathcal{P}^1$ (see \eqref{eq:parents}), which in turn may be receiving messages from other sensors. Thus, sensor $1$'s decision signals information about its own observations as well as the information it gathers  due to signaling by sensors in the set  $\mathcal{P}^1$.


Some basic questions associated with signaling problems with the above features are: What is an information state (sufficient statistic)  for the sensors? How is signaling incorporated in evolution/update of the information state? Is there an explicit description of all signaling possibilities?  
 We will answer these questions in Sections \ref{sec:IS} -  \ref{sec:thresholds} and  discuss them further in Section \ref{sec:con}.

\section{Information States} \label{sec:IS}

In this section, we identify information states for the  sensors. We start by fixing the strategies of all sensors~$j \neq i$ to an arbitrary choice and considering the problem of minimizing the expected cost only over the strategy of sensor $i$. When the strategies of all other sensors are fixed, we show that sensor $i$ can optimally make its decisions as a function of its posterior belief on the hypothesis and its received messages. Therefore, the posterior belief and the received messages constitute an \emph{information state} for sensor $i$ when strategies of all other sensors are fixed.  This result does not depend on the arbitrary choice of other sensors' strategies. In particular, if all other sensors were using their globally optimal strategies, sensor $i$'s optimal strategy would still be a function of its information state. 

\begin{definition}   
  Given fixed  strategies for all sensors $j \neq i$ and decision rules $\gamma^i_{1:t-1}$ for sensor $i$, we define sensor~$i$'s belief on the hypothesis given all its information at time $t$ as 
  \[ \Pi^i_t := \mathds{P}(H=0|Y^i_{1:t},  U^{\mathcal{P}^i}_{1:t-1}, U^i_{1:t-1}=b_{1:t-1}), \] 
 where $b_{1:t-1}$ denotes a sequence of blank messages from time $1$ to $t-1$. 
 For \(t=0\), we define \(\Pi^i_0 := p_0\).
    
\end{definition}

Note that the belief $\Pi^i_t$ is a random variable whose realizations (denoted by $\pi^i_t$) depend on the realizations of observations and messages received by sensor $i$. Also note that we define $\Pi^i_t$ assuming that sensor $i$ has not stopped before time $t$, that is, $U^i_{1:t-1} =b_{1:t-1}$. If the sensor has stopped before time $t$, it does not have to make any decision at time $t$ and therefore it would be meaningless to define its information state.
 
 We now describe the evolution of  $\Pi^i_t$  in the following lemma.
\medskip
\begin{lemma}\label{lemma:update}
Sensor $i$'s belief $\Pi^i_t$ evolves according to the following equation
\begin{equation}
\Pi^i_{t+1} = \eta^i_t(\Pi^i_t, Y^i_{t+1}, U^{\mathcal{P}^i}_{1:t}),
\end{equation}
where $\eta^i_t$ is a deterministic function that depends on other sensors' strategies.
%
\end{lemma}   
\begin{proof} See Appendix \ref{sec:lemma_1}.
 \end{proof}

The optimal strategy for sensor~i (for the given choice of $\Gamma^j, j \neq i$) can be obtained by means of a dynamic program. We  define below the value functions of this dynamic program.

\begin{definition}\label{def:value_functions}
\begin{enumerate}[(i)]
\item For each realization $\pi^i_T, u^{\mathcal{P}^i}_{1:T-1}$ of $\Pi^i_T, U^{\mathcal{P}^i}_{1:T-1}$, we define

\begin{align}
   &V^i_T(\pi^i_T, u^{\mathcal{P}^i}_{1:T-1}) := \min \Big\{  \mathds{E}[ J(H,\{U^i_{\tau^i}, \tau^i\}_{i \in \mathcal{N}})|\pi^i_T, \notag \\
   &   u^{\mathcal{P}^i}_{1:T-1}, U^i_{1:T-1} = b_{1:T-1}, U^i_T=0], \nonumber \\
  	                                 &\mathds{E}[ J(H,\{U^i_{\tau^i}, \tau^i\}_{i \in \mathcal{N}})|\pi^i_T, u^{\mathcal{P}^i}_{1:T-1}, U^i_{1:T-1} = b_{1:T-1},  \notag \\&~~~~~~U^i_T=1], \Big \}\label{eq:dpeq1}
\end{align}

\item For $t = T-1,\ldots, 1,$ and for each realization $\pi^i_t, u^{\mathcal{P}^i}_{1:t-1}$ of $\Pi^i_t, U^{\mathcal{P}^i}_{1:t-1}$, we define
\begin{align}
&V^i_t(\pi^i_t, u^{\mathcal{P}^i}_{1:t-1}) := \min \Big\{ \mathds{E}[ J(H,\{U^i_{\tau^i}, \tau^i\}_{i \in \mathcal{N}})|\pi^i_t, \notag \\
   &  u^{\mathcal{P}^i}_{1:t-1}, U^i_{1:t-1} = b_{1:t-1}, U^i_t=0], \nonumber \\
  	                                 &\mathds{E}[ J(H,\{U^i_{\tau^i}, \tau^i\}_{i \in \mathcal{N}})|\pi^1_t,  u^{\mathcal{P}^i}_{1:t-1}, U^i_{1:t-1} = b_{1:t-1}, \notag \\&~~~~~~ U^i_t=1], \nonumber \\
  	                         &  \mathds{E}[V^i_{t+1}(\Pi_{t+1},u^{\mathcal{P}^i}_{1:t-1},U^{\mathcal{P}^i}_{t} )|\pi^1_t, u^{\mathcal{P}^i}_{1:t-1}, U^i_{1:t-1} = b_{1:t-1}, \notag \\&~~~~~U^i_t=b ] \Big \}     \label{eq:dpeq2} 
 \end{align}
\end{enumerate}
\end{definition}       
\medskip
\begin{theorem}\label{thm:DP}
  With  fixed strategies for sensors~$j \neq i$, there is an optimal strategy for sensor~$i$ of the form:
  \begin{equation*} \label{eq:Qp1}
  U^i_t = \gamma^i_{t}(\Pi^{i}_{t}, U^{\mathcal{P}^i}_{1:t-1})
  \end{equation*}
  for \(t=1,2,...,T\). Moreover, this optimal strategy can be obtained by the dynamic program described by the value functions in Definition~\ref{def:value_functions}. At time $t$ and for a given $\pi^i_t$ and $u^{\mathcal{P}^i}_{1:t-1}$, the optimal decision is $0$ (or $1$/$b$) if the first (or second/third) term is the minimum in the definition of $V^i_t(\pi^i_t, u^{\mathcal{P}^i}_{1:t-1})$.
\end{theorem}
\begin{proof} See Appendix \ref{sec:info_states_proof}.
\end{proof}

\section{A counterexample to two threshold rules}\label{sec:counter}
In the sequential detection problem with a single sensor \cite{Wald}, it is well known that an optimal strategy is a function of the sensor's posterior belief \(\Pi_t\) and is described by two thresholds at each time. That is, the decision at time \(t\), \(Z_t\), is given as:
      \[ Z_t = \left \{ \begin{array}{ll}
               1 & \mbox{if $\Pi_t \leq \alpha_t$} \\
               b & \mbox{if $\alpha_t<\Pi_t < \beta_t$} \\
               0 & \mbox{if $\Pi_t \geq \beta_t$}
               \end{array}
               \right. \]
     where \(b\) denotes a decision to continue taking observations and \(\alpha_t \leq \beta_t\) are real numbers in \([0,1]\).
 A similar two-threshold structure of optimal strategies was also established for the decentralized Wald problem in \cite{Dec_Wald}. We will show by means of a counterexample that such a structure is not necessarily optimal in our problem.
 
 Consider the following instance of our problem\footnote{This example is a modification of an example presented  in \cite{Nayyar_MTNS}.}. There are two sensors and we have equal prior on \(H\), that is \[\mathds{P}(H=0)=\mathds{P}(H=1)=1/2\] and a time horizon of \(T=3\). The operational costs are given as
 \begin{equation}
 O(\tau^1,\tau^2) = \sum_{t=1}^T c(\mathcal{A}_t),
 \end{equation} 
 where  $c(\emptyset)=0$, $c(\{1\}) = c(\{2\})=1$ and $c(\{1,2\}) = K$, $1<K<2$ (recall that $\mathcal{A}_t$ is the set of active sensors at time $t$, see \eqref{eq:active}). The observation space of sensor $1$ is \(\mathcal{Y}^1 = \{0,1\}\) and the observations at time \(t\) obey the following conditional probabilities: 
\begin{center}
\begin{tabular}{|l|c|c|}
\hline
Observation, $y$ & $y=0$ & $y=1$ \\
\hline
$f^1_t(y|H=0)$    & $q_t$ & $(1-q_t)$  \\
\hline
$f^1_t(y|H=1)$    & $(1-q_t)$ & $q_t$  \\
\hline
\end{tabular}
\end{center}
where $q_1 =q_2 =1/2$ and $q_3=1$. Thus, the first two observations of sensor $1$ reveal no information about $H$ while the third observation reveals $H$ noiselessly.
The observation space of sensor $2$ is \(\mathcal{Y}^2 = \{0,1,2\}\) and the observations at time \(t\) obey the following conditional probabilities:
\begin{center}
\begin{tabular}{|l|c|c|c|}
\hline
Observation, $y$ & $y=0$ & $y=1$ & $y=2$\\
\hline
$f^1_t(y|H=0)$    & $r_t$ & $(1-r_t)$ & $0$  \\
\hline
$f^1_t(y|H=1)$    & $0$ &$(1-r_t)$ & $r_t$  \\
\hline
\end{tabular}
\end{center}
where \(r_2=r_3=0\) and $0<r_1<1$. Thus, the second and third observations of sensor $2$ reveal no information about $H$. 

Both sensors can receive each other's final decision, that is, $\mathcal{P}^1 = \{2\}$ and $\mathcal{P}^2 = \{1\}$. The terminal cost function is given as 
\begin{align}
A(H,U^1_{\tau^1}, U^2_{\tau^2},\tau^1,\tau^2) = \left\{ \begin{array}{ll}
                                                       a(H, U^1_{\tau^1}) & \mbox{if $\tau^1 \geq \tau^2$} \\
                                                       a(H, U^2_{\tau^2}) & \mbox{if $\tau^2 > \tau^1$}
                                                       \end{array}
                                                    \right.
\end{align}
where $a(x,y) := \mu\mathds{1}_{\{x \neq y\}}$ and $\mu >0$. This cost structure implies that the final decision of the sensor that stops later is taken as the system decision about the hypothesis. The constant $\mu$ can be interpreted as the cost of making a mistake in the system decision.

Note that under this statistical model of observations, there exists a choice of strategies such that the system makes perfect final decision on the hypothesis and incurs only operational costs (if sensor $2$ stops at $t=1$ and sensor $1$ waits till time $t=3$, then it can make a perfect decision on $H$ and the system incurs an operational cost of $K+2$). 
We assume that the cost of  a mistake in the system decision (that is, the parameter $\mu$)  is sufficiently high so that any choice of strategies that makes a mistake in the system decision with non-zero probability will have a performance worse than $K+2$. Thus, any choice of strategies that makes a mistake in the final decision with non-zero probability cannot be optimal.
\par

In the above instance of our problem, sensor $2$'s posterior belief  on the event $\{H=0\}$ after making the observation at time $t=1$, $\Pi^2_1$, can take only one of three values - $0,1/2$ or $1$. If sensor $2$ is restricted to use a two-threshold rule at time $t=1$, then the lowest achievable value of the objective is given as:
\begin{align} \label{eq:threshold_cost}
\min\big[\{K + r_1 + (1-r_1)(K+1)\}, \{K +2-r_1/2\}\big]. 
\end{align}
The first term in the minimization in \eqref{eq:threshold_cost} corresponds to the case when $\gamma^2_1$ is given as:
\begin{equation} \label{eq:ex_1}
 U^2_1 = \left \{ \begin{array}{ll}
               1 & \mbox{if $\Pi^2_1 = 0$} \\
               b & \mbox{if $0<\Pi^2_1<1$} \\
               0 & \mbox{if $\Pi^2_1 = 1$}
               \end{array}
               \right. ,\end{equation}
               $\gamma^2_2$ is such that sensor $2$ stops at time $t=2$ and sensor $1$'s strategy is as follows: if it receives a $0$ or $1$ from sensor $2$ at time $t=1$, it stops at time $t=2$ and declares the received message as its final decision, otherwise it continues operating till time $t=3$ when it observes $H$ noiselessly and declares the observed $H$ as its decision. The
  the second term in the minimization in \eqref{eq:threshold_cost} corresponds to $\gamma^2_1$ being
\begin{equation} \label{eq:ex_2}  
 U^2_1 = \left \{ \begin{array}{ll}
               1 & \mbox{if $\Pi^2_1 < 1$} \\
               0 & \mbox{if $\Pi^2_1 = 1$}
               \end{array}
               \right. ;\end{equation}
 sensor $1$'s strategy is as follows: if it receives a $0$ from sensor $2$ at time $t=1$, it stops at time $t=2$ and declares the received message as its final decision, otherwise it continues operating till time $t=3$ when it  declares the observed $H$ as its decision. It can be easily verified that other choices of thresholds for sensor~2 at time $t=1$ do not give a lower value than the expression in (\ref{eq:threshold_cost}).

 Consider now the following choice of $\gamma^{2,*}_1$:
 \begin{equation}\label{eq:non_th}
  U^2_1 = \left \{ \begin{array}{ll}
               1 & \mbox{if $\Pi^2_1 = 0$} \\
               0 & \mbox{if $0<\Pi^2_1<1$} \\
               b & \mbox{if $\Pi^2_1 = 1$}
               \end{array}
               \right. ,\end{equation}
                 $\gamma^2_2$ is such that sensor $2$ stops and decides $U^2_2 =0$ at time $t=2$  and sensor $1$'s strategy is as follows: if it receives a $1$ from sensor $2$ at time $t=1$, it stops at time $t=2$ and declares the received message as its final decision, if it receives $b$ from sensor $2$, it stops at time $t=2$ and declares $U^1_2=0$, otherwise it continues operating till time $t=3$ when it  declares the observed $H$ as its decision.
 The  expected cost in this case is $\mathcal{J}^{*} = K+ 2(1-r_1) + r_1(K+1)/2$. It is easy to check that for $1<K<2$ and $r_1<2/3$,
 \[ \mathcal{J}^{*} < K+\min[\{r_1 + (1-r_1)(K+1)\}, \{2-r_1/2\}]\]
 Thus, $\gamma^{2,*}_1$ outperforms the two-threshold rules.  
 \medskip
 
 \emph{Discussion:} The threshold rule of \eqref{eq:ex_1} implies that a message of $0$ or $1$ from sensor $2$ at time $t=1$ conveys certainty about the hypothesis. On receiving either $0$ or $1$ at time $t=1$, sensor $1$ can declare the correct hypothesis and stop at time $t=2$. However, if sensor $2$ is uncertain at time $t=1$ (that is, $0 < \Pi^2_1 <1$), then it does not stop at time $t=1$ and incurs additional operation costs at time $t=2$ even though its observation at time $t=2$ provides no information. By making the probability $r_1$ small, the contribution of these wasteful operational costs in the overall system cost is increased. The threshold rule of \eqref{eq:ex_2} ensures that sensor $2$  incurs no unnecessary operational costs by always stopping at time $t=1$. However, in doing so, it reduces the information content of the message sent by sensor $2$ at time $t=1$ (since $U^2_1=1$ no longer conveys certainty about $H$). 
 
 On the other hand, the non-threshold rule of \eqref{eq:non_th} attempts to minimize sensor $2$'s operational costs by making sure that it stops at time $1$ if it is certain that $H=1$ or if it is uncertain. A decision not to stop at time $t=1$ by sensor $2$ \emph{signals} to sensor $1$ that sensor $2$ is certain that $H=0$. In that case, both sensors can stop at $t=2$ and declare the correct hypothesis. The cost and observation parameters are chosen so that this signaling strategy outperforms the two-threshold based strategies.

\section{Parametric Characterization of Optimal Strategies} \label{sec:thresholds}

In centralized sequential detection problems, the two threshold characterization of optimal strategies modifies the problem of finding the globally optimal strategies from a sequential functional optimization problem to a sequential parametric optimization problem. Even though we have established that a classical two threshold rule does not hold for our detection problem, it is still possible to get a finite parametric characterization of optimal strategies. Such a parametric characterization provides significant computational advantage in finding optimal strategies by reducing the search space for an optimal strategy.        

 In Theorem~\ref{thm:DP}, we have established  that for an arbitrarily fixed choice of other sensors' strategies, the optimal strategy for sensor $i$ can be determined by a dynamic program using the value functions \(V^i_t(\pi,u^{\mathcal{P}^i}_{1:t-1}), t=T,...,2,1\). 
 We now establish the following lemma about these value functions.
 \begin{lemma} \label{lemma:lemma_2}
  With  fixed strategies for sensors~$j \neq i$, the value function $V^i_T$  can be expressed as:
  \begin{equation}
  V^i_{T}(\pi,u^{\mathcal{P}^i}_{1:T-1}) := \min\{ l^{0}(\pi,u^{\mathcal{P}^i}_{1:T-1}), l^{1}(\pi,u^{\mathcal{P}^i}_{1:T-1})\} 
  \end{equation}
  where for each $u^{\mathcal{P}^i}_{1:T-1}$, $l^{0}(\cdot, u^{\mathcal{P}^i}_{1:T-1})$ and $l^{1}(\cdot,u^{\mathcal{P}^i}_{1:T-1})$ are affine functions of $\pi$.
  
  Also, the value functions at time \(t\) can be expressed as:
  \begin{align}
  V^i_{t}(\pi, u^{\mathcal{P}^i}_{1:t-1}) := \min\{ &L^{0}_t(\pi,u^{\mathcal{P}^i}_{1:t-1}), L^{1}_t(\pi,u^{\mathcal{P}^i}_{1:t-1}), \notag \\ &G_t(\pi,u^{\mathcal{P}^i}_{1:t-1})\}\label{eq:thresholds_lemma1}
  \end{align}
   where for each $u^{\mathcal{P}^i}_{1:t-1}$, $L^{0}(\cdot, u^{\mathcal{P}^i}_{1:t-1})$ and $L^{1}_t(\cdot,u^{\mathcal{P}^i}_{1:t-1})$ are affine functions of $\pi$ and $G_t(\cdot,u^{\mathcal{P}^i}_{1:t-1})$ is a concave function of $\pi$.
   \end{lemma}
 \begin{proof} See Appendix \ref{sec:lemma_2}. \end{proof}
 
 We can now derive the following result from the above lemma.
 \begin{theorem}\label{thm:thresholds}
  With  fixed strategies for sensors~$j \neq i$,  an optimal strategy for sensor $i$ can be characterized as follows: For each time $t$ and each realization of messages $u^{\mathcal{P}^i}_{1:t-1}$ received by sensor $i$, there exist sub-intervals $[\alpha_t(u^{\mathcal{P}^i}_{1:t-1}),\beta_t(u^{\mathcal{P}^i}_{1:t-1})]$ and $[\delta_t(u^{\mathcal{P}^i}_{1:t-1}),\theta_t(u^{\mathcal{P}^i}_{1:t-1})]$ of  $[0,1]$ such that 
   
  \begin{align}
  U^i_t = \left \{ \begin{array}{ll}
                       1 & \mbox{if $\Pi^i_t \in [\alpha_t(u^{\mathcal{P}^i}_{1:t-1}) ,\beta_t(u^{\mathcal{P}^i}_{1:t-1})]$} \\
                       0  & \mbox{if $ \Pi^i_t \in [\delta_t(u^{\mathcal{P}^i}_{1:t-1}) ,\theta_t(u^{\mathcal{P}^i}_{1:t-1})]$} \\                       
               b & \mbox{otherwise}                        
               \end{array}
               \right.  ,
                 \end{align}
                 For $t=T$, \[ [\alpha_T(u^{\mathcal{P}^i}_{1:T-1}) ,\beta_T(u^{\mathcal{P}^i}_{1:T-1})] \cup [\delta_T(u^{\mathcal{P}^i}_{1:T-1}) ,\theta_T(u^{\mathcal{P}^i}_{1:T-1})] = [0,1].\]

\end{theorem}
\begin{proof}
From Lemma~\ref{lemma:lemma_2}, we know that the value functions for $t<T$ can be written as minimum of two affine and one concave functions. The intervals correspond to regions where one of the affine functions coincides with the minimum.  At time $T$, the value function is the minimum of two affine functions which implies that the two intervals cover $[0,1]$.
\end{proof}

The result of Theorem \ref{thm:thresholds} provides a 4 - threshold characterization of optimal decision strategy for a sensor where the thresholds depend on the received messages. 

If sensor $i$'s decision at its stopping time is not interpreted as a decision about the hypothesis $H$ but simply as a message to other sensors, then it may be allowed to take more than binary values, that is, if $U^i_t \neq b$, then it may take values in $\{0,1,\ldots,M-1\}$. The arguments of Theorems \ref{thm:DP} and \ref{thm:thresholds} can be easily extended to obtain the following result.

\begin{corollary}
If sensor $i$ can send one of $M$ possible messages when it decides to stop, then its optimal decision strategy is characterized by $M$ sub-intervals of $[0,1]$: $[\alpha^0_t(u^{\mathcal{P}^i}_{1:t-1}) ,\beta^0_t(u^{\mathcal{P}^i}_{1:t-1})], \ldots, [\alpha^{M-1}_t(u^{\mathcal{P}^i}_{1:t-1}) ,\beta^{M-1}_t(u^{\mathcal{P}^i}_{1:t-1})]$ such that the optimal decision is $U^i_t = m$ if $\Pi^i_t \in [\alpha^m_t(u^{\mathcal{P}^i}_{1:t-1}) ,\beta^m_t(u^{\mathcal{P}^i}_{1:t-1})]$.
\end{corollary}

\section{Discussion and Conclusions}\label{sec:con}

We considered a sequential detection problem in sensor networks where each sensor can communicate with only a subset of sensors in the network, sensors cannot communicate continuously, each sensor makes its own stopping decision and communicates only once after making the decision to stop. Even though communication is not continuous, the absence of communication conveys information. Sensor $j$'s decisions $u^j_{1:t}$ signal to other sensors in the set $\mathcal{C}^j$ the information sensor $j$ has up to time $t$ through its own observations $Y^j_{1:t}$ and the messages $U^{\mathcal{P}^j}_{1:t-1}$ it received from sensors in the set $\mathcal{P}^j$. Since sensors in $\mathcal{P}^j$ may be receiving messages from  sensors with which sensor $j$ has no direct communication, the messages $U^{\mathcal{P}^j}_{1:t-1}$  include information about (some) sensors not in $\mathcal{P}^j$. This form of signaling that arises from the decentralization of stopping decisions as well as the fact that sensors transmit their decisions only locally is not present in static detection problems or in sequential detection problems where stopping decisions are centralized. Some of the basic questions associated with signaling problems with the above features were posed in \ref{sec:PF} and were analyzed and answered in \ref{sec:IS}-\ref{sec:thresholds}. Here we present a qualitative discussion of the answers to these questions.

We have shown in Section \ref{sec:IS} that a sensor's posterior belief on the hypothesis conditioned on its observations and the messages received from other sensors is an appropriate information state for the sensor. Signaling is done based on this information state, and all signaling possibilities are captured by a four threshold decision rule where the thresholds depend on received messages. In determining its signaling action at any time $t$ a sensor must assess the effect of actions of all other sensors on the cost. This assessment is described by expressions such as those in \eqref{eq:ap2.2} and \eqref{eq:ap2.4} (in Appendix~\ref{sec:info_states_proof}). The effect of signaling is also  explicitly taken into account in the update of sensor's information state; it is described in Step 2 of the proof of Lemma \ref{lemma:update}. As pointed out earlier in this section, the messages sensor $j$ receives from other sensors in $\mathcal{P}^j$ convey information about the observations of sensors in $\mathcal{N}\setminus\mathcal{P}^j$; this communication of information is described in Step 2 of the proof of Lemma~\ref{lemma:update}.

The results of this paper reveal qualitative properties of optimal sensor signaling strategies. Moreover, if the strategies $\Gamma^{-i} :=(\Gamma^1,\Gamma^2,\ldots,\Gamma^{i-1},\Gamma^{i+1},\ldots,\Gamma^N)$ of all sensors other than sensor $i$ have already been chosen, Theorem \ref{thm:DP} provides a dynamic programming method to find the best strategy $\Gamma^i$ of sensor $i$ for the given choice of $\Gamma^{-i}$. An iterative application of such an approach may be used to find person-by-person optimal strategies \cite{Ho1980}. Finding globally optimal strategies for the problem formulated in this paper is a difficult task since it involves non-convex functional optimization \cite{Ho1980}. 

\appendices
\section{Proof of Lemma~\ref{lemma:update}} \label{sec:lemma_1}
\begin{proof}
We will prove the lemma for $i=1$. We will proceed in three steps.


\textbf{Step 1:} Consider a realization $y^1_{1:t}$ of sensor $1$'s  observations and a realization $u^{\mathcal{P}^1}_{1:t-1}$ of the messages received by sensor $i$ before it makes its decision at time $t$. By definition, the realization $\pi^1_t$ of sensor $1$'s belief $\Pi^1_t$ is
\begin{equation}
\pi^1_t = \mathds{P}(H=0|y^1_{1:t},u^{\mathcal{P}^1}_{1:t-1},u^1_{1:t-1}=b_{1:t-1}) 
\end{equation}
In the first step of the proof, we will use $\pi^1_t$ and $u^{\mathcal{P}^1}_{1:t-1}$ to construct sensor $1$'s joint belief on $H$ \emph{and} the observations of all other sensors. Recall that the strategies $\Gamma^j, j\neq 1$ have been fixed. Under these fixed strategies, $U^{\mathcal{P}^1}_{1:t} $ is a deterministic function of $Y^{2:N}_{1:t}$ and $U^1_{1:t-1}$.  To see this, think of the group of sensors $2$ to $N$ as a single deterministic system interacting with sensor $1$.  Until time $t$, the inputs to this system are the observations $Y^{2:N}_{1:t}$ and the messages from sensor $1$, $U^1_{1:t-1}$. The outputs of this system, as seen by sensor $1$, are $U^{\mathcal{P}^i}_{1:t}$. With the strategies of all sensors $2$ to $N$ fixed, this is a deterministic system with a fixed deterministic mapping from its inputs $Y^{2:N}_{1:t}$ and $U^1_{1:t-1}$ to its outputs $U^{\mathcal{P}^i}_{1:t}$.  Let this mapping be denoted by $\mathcal{M}_t$, that is,
\begin{equation}\label{eq:sys_eq1}
U^{\mathcal{P}^1}_{1:t} = \mathcal{M}_t(Y^{2:N}_{1:t},U^1_{1:t-1})
\end{equation}
Similarly, for a fixed choice of $\gamma^1_{1:t-1}$, $U^1_{1:t-1}$ is a deterministic function of $Y^1_{1:t-1}, U^{\mathcal{P}^1}_{1:t-2}$,
\begin{equation}\label{eq:sys_eq2}
 U^1_{1:t-1} = \mathcal{N}_t(Y^{1}_{1:t-1},U^{\mathcal{P}^1}_{1:t-2})
\end{equation}

Now, let $\rho^1_t(h, y^{2:N}_{1:t})$ be defined as sensor $1$'s joint belief at time $t$ on $H$ and the observations of all other sensors. That is,
\begin{align}
&\rho^1_t(h,y^{2:N}_{1:t}) \notag \\ &:= \mathds{P}(H=h, Y^{2:N}_{1:t} = y^{2:N}_{1:t}|y^1_{1:t},u^{\mathcal{P}^1}_{1:t-1},u^1_{1:t-1}=b_{1:t-1}) 
\end{align}
For $h=0$, we can write $\rho^1_t(0,y^{2:N}_{1:t})$ as
\begin{align}
& \mathds{P}( Y^{2:N}_{1:t} = y^{2:N}_{1:t}|H=0,y^1_{1:t},u^{\mathcal{P}^1}_{1:t-1},u^1_{1:t-1}=b_{1:t-1}) \notag \\&\times \mathds{P}(H=0|y^1_{1:t},u^{\mathcal{P}^1}_{1:t-1},u^1_{1:t-1}=b_{1:t-1}) \notag\\
&= \mathds{P}( Y^{2:N}_{1:t} = y^{2:N}_{1:t}|H=0,y^1_{1:t},u^{\mathcal{P}^1}_{1:t-1},u^1_{1:t-1}=b_{1:t-1})\pi^1_t \label{eq:ap1a}
\end{align}

Using Bayes' rule, the first term of \eqref{eq:ap1a} can be written as 
\begin{align}
 \frac{\mathds{P}(H=0, y^{2:N}_{1:t},  u^{\mathcal{P}^1}_{1:t-1},U^1_{1:t-1}=b_{1:t-1}|y^1_{1:t} ) }{\sum_{ \tilde{y}^{2:N}_{1:t}}\mathds{P}(H=0, \tilde{y}^{2:N}_{1:t}, u^{\mathcal{P}^1}_{1:t-1},U^1_{1:t-1}=b_{1:t-1}|y^1_{1:t})}  \label{eq:ap1.0}
\end{align}
Because of \eqref{eq:sys_eq1} and \eqref{eq:sys_eq2}, the numerator in \eqref{eq:ap1.0} can be written as
\begin{align}
&\mathds{1}_{\{u^{\mathcal{P}^1}_{1:t-1} =\mathcal{M}_t(y^{2:N}_{1:t},b_{1:t-1}) \}} \mathds{1}_{\{(b_{1:t-1}) =\mathcal{N}_t(y^1_{1:t-1}, u^{\mathcal{P}^1}_{1:t-2}) \}} \notag \\
&\times \mathds{P}(H=0,  y^{2:N}_{1:t}|y^1_{1:t} ) \notag \\
&=\mathds{1}_{\{u^{\mathcal{P}^1}_{1:t-1} =\mathcal{M}_t(y^{2:N}_{1:t},b_{1:t-1}) \}} \mathds{1}_{\{(b_{1:t-1}) =\mathcal{N}_t(y^1_{1:t-1}, u^{\mathcal{P}^1}_{1:t-2}) \}} \notag \\
&\times \mathds{P}( y^{2:N}_{1:t}|H=0,y^1_{1:t} ) \mathds{P}(H=0|y^1_{1:t}) \notag \\
&=\mathds{1}_{\{u^{\mathcal{P}^1}_{1:t-1} =\mathcal{M}_t(y^{2:N}_{1:t},b_{1:t-1}) \}} \mathds{1}_{\{(b_{1:t-1}) =\mathcal{N}_t(y^1_{1:t-1}, u^{\mathcal{P}^1}_{1:t-2}) \}} \notag \\
&\times \mathds{P}( y^{2:N}_{1:t}|H=0 ) \mathds{P}(H=0|y^1_{1:t}), \label{eq:ap1.8}
\end{align}
where we used the conditional independence of observations in \eqref{eq:ap1.8}.
Similar expressions can be obtained for each term in the denominator of \eqref{eq:ap1.0}. Substituting these expressions back in \eqref{eq:ap1.0} and canceling the common terms simplifies \eqref{eq:ap1.0} to
\begin{align}
\frac{\mathds{1}_{\{u^{\mathcal{P}^1}_{1:t-1} =\mathcal{M}_t(y^{2:N}_{1:t},b_{1:t-1}) \}}\mathds{P}( y^{2:N}_{1:t}|H=0 )}{\sum_{ \tilde{y}^{2:N}_{1:t}}\mathds{1}_{\{u^{\mathcal{P}^1}_{1:t-1} =\mathcal{M}_t(\tilde{y}^{2:N}_{1:t},b_{1:t-1}) \}}\mathds{P}( \tilde{y}^{2:N}_{1:t}|H=0 )} \label{eq:ap1.2}
\end{align}
Substituting \eqref{eq:ap1.2} back in \eqref{eq:ap1a} gives
\begin{align}
&\rho^1_t(0,y^{2:N}_{1:t}) \notag \\&= \frac{\mathds{1}_{\{u^{\mathcal{P}^1}_{1:t-1} =\mathcal{M}_t(y^{2:N}_{1:t},b_{1:t-1}) \}}\mathds{P}( y^{2:N}_{1:t}|H=0 )}{\sum_{ \tilde{y}^{2:N}_{1:t}}\mathds{1}_{\{u^{\mathcal{P}^1}_{1:t-1} =\mathcal{M}_t(\tilde{y}^{2:N}_{1:t},b_{1:t-1}) \}}\mathds{P}( \tilde{y}^{2:N}_{1:t}|H=0 )}  \pi^1_t \label{eq:ap1.6}
\end{align}

Similarly, we can get
\begin{align}
&\rho^1_t(1,y^{2:N}_{1:t}) =\notag \\& \frac{\mathds{1}_{\{u^{\mathcal{P}^1}_{1:t-1} =\mathcal{M}_t(y^{2:N}_{1:t},b_{1:t-1}) \}}\mathds{P}( y^{2:N}_{1:t}|H=1 )}{\sum_{ \tilde{y}^{2:N}_{1:t}}\mathds{1}_{\{u^{\mathcal{P}^1}_{1:t-1} =\mathcal{M}_t(\tilde{y}^{2:N}_{1:t},b_{1:t-1}) \}}\mathds{P}( \tilde{y}^{2:N}_{1:t}|H=1 )}  (1-\pi^1_t) \label{eq:ap1.7}
\end{align}

 \textbf{Step 2:} We now consider how sensor $1$ can update its belief on $H$ and $Y^{2:N}_{1:t}$ after observing the messages  $u^{\mathcal{P}^1}_t$  at time $t$. 
 We define
 \begin{align}
 \sigma^1_t(h,y^{2:N}_{1:t}) := \mathds{P}(H=h,  y^{2:N}_{1:t}|y^1_{1:t},u^{\mathcal{P}^1}_{1:t},u^1_{1:t}=b_{1:t}),
 \end{align}
 which using Bayes' rule can be written as
 \begin{align}
 \frac{\mathds{P}(H=h, Y^{2:N}_{1:t} = y^{2:N}_{1:t}, u^{\mathcal{P}^1}_{t}|y^1_{1:t},u^1_{1:t}=b_{1:t}, u^{\mathcal{P}^1}_{1:t-1})}{\sum_{\tilde{h},\tilde{y}^{2:N}_{1:t}}\mathds{P}(H=\tilde{h}, Y^{2:N}_{1:t} = \tilde{y}^{2:N}_{1:t}, u^{\mathcal{P}^1}_{t}|y^1_{1:t},u^{\mathcal{P}^1}_{1:t-1},u^1_{1:t}=b_{1:t}) } \label{eq:ap1.3}
 \end{align}
 Consider the numerator in \eqref{eq:ap1.3}. It can be written as
 \begin{align}
 \mathds{P}(H=h, Y^{2:N}_{1:t} = y^{2:N}_{1:t}, u^{\mathcal{P}^1}_{t}|y^1_{1:t},u^{\mathcal{P}^1}_{1:t-1},u^1_{1:t-1}=b_{1:t-1}) \label{eq:ap1.4}
 \end{align}
 where we dropped the conditioning on $u^1_t$ because $u^1_t$ is a function of the rest of terms included in the conditioning. For a fixed choice of other sensors' strategies, $u^{\mathcal{P}^1}_t$ is a deterministic fixed function of $Y^{2:N}_{1:t}$ and $U^1_{1:t-1}$. Let this map be $\mathcal{L}_t$, that is,
 \begin{equation}
U^{\mathcal{P}^1}_{t}= \mathcal{L}_t(Y^{2:N}_{1:t},U^1_{1:t-1}) \label{eq:Lt}
\end{equation}
We can now write \eqref{eq:ap1.4} as
\begin{align}
&\mathds{1}_{\{u^{\mathcal{P}^1}_t =\mathcal{L}_t(y^{2:N}_{1:t},b_{1:t-1}) \}} \notag \\&\times \mathds{P}(H=h,  y^{2:N}_{1:t}|y^1_{1:t}, u^{\mathcal{P}^1}_{1:t-1}, u^1_{1:t-1}=b_{1:t-1}) \notag \\
&= \mathds{1}_{\{u^{\mathcal{P}^1}_t) =\mathcal{L}_t(y^{2:N}_{1:t},b_{1:t-1})\} }\times 
\rho^1_t(h,y^{2:N}_{1:t}) \label{eq:ap1.5}
\end{align}
 Substituting \eqref{eq:ap1.5} in \eqref{eq:ap1.3}, we obtain that 
 \begin{align}
 &\sigma^1_t(h,y^{2:N}_{1:t}) = \notag \\&\frac{\mathds{1}_{\{u^{\mathcal{P}^1}_t =\mathcal{L}_t(y^{2:N}_{1:t},b_{1:t-1}) \} }\times 
\rho^1_t(h,y^{2:N}_{1:t})}{\sum_{\tilde{h},\tilde{y}^{2:N}_{1:t}} \mathds{1}_{\{u^{\mathcal{P}^1}_t =\mathcal{L}_t(\tilde{y}^{2:N}_{1:t},b_{1:t-1}) \} }\times 
\rho^1_t(\tilde{h},\tilde{y}^{2:N}_{1:t})} \label{eq:ap1.10}
 \end{align}
 We let $\sigma^1_t(h) = \sum_{y^{2:N}_{1:t}} \sigma^1_t(h,y^{2:N}_{1:t})$.

 \textbf{Step 3:} Finally, we will show that $\pi^1_{t+1}$ can be obtained from $\sigma^1_t(\cdot)$ and $y^1_{t+1}$.
 \begin{align}
& \pi^1_{t+1} = \mathds{P}(H=0|y^1_{1:t+1}, u^{\mathcal{P}^1}_{1:t-1}, U^1_{1:t-1}=b_{1:t-1}) \notag \\
 &= \frac{\mathds{P}( y^1_{t+1},H=0|y^1_{1:t}, u^{\mathcal{P}^1}_{1:t-1}, U^1_{1:t-1}=b_{1:t-1})}{\sum_{h} \mathds{P}(y^1_{t+1},H=h|y^1_{1:t}, u^{\mathcal{P}^1}_{1:t-1}, U^1_{1:t-1}=b_{1:t-1}) } \notag\\
 &= \frac{\mathds{P}(y^1_{t+1}|H=0)\sigma^1_t(0)}{\mathds{P}(y^1_{t+1}|H=0)\sigma^1_t(0) + \mathds{P}(Y^1_{t+1}=y^1_{t+1}|H=1)\sigma^1_t(1)} \notag \\
  &= \frac{f^1_{t+1}(y^1_{t+1}|0)\sigma^1_t(0)}{f^1_{t+1}(y^1_{t+1}|0)\sigma^1_t(0) + f^1_{t+1}(y^1_{t+1}|1)\sigma^1_t(1)} \label{eq:ap1.9}
 \end{align}
~\\
Combining \eqref{eq:ap1.9}, \eqref{eq:ap1.10}, \eqref{eq:ap1.6} and \eqref{eq:ap1.7}, it follows that $\Pi^1_{t+1} = \eta^1_t(\Pi^1_t, Y^1_{t+1}, U^{\mathcal{P}^1}_{1:t})$.

\end{proof}

\section{Proof of Theorem \ref{thm:DP}} \label{sec:info_states_proof}

We will prove Theorem \ref{thm:DP} for $i=1$ by a backward inductive argument. Suppose that the sensor $1$ is still active at the terminal time $T$. Let  $\mathcal{F}^1_T$ be the set of sensors in $\mathcal{P}^1$ that have stopped before time $T$, that is, $\mathcal{F}^1_T = \mathcal{P}^1\setminus\mathcal{A}_T$ and let $\mathcal{G}^1_T = \mathcal{N} \setminus(\{ 1\} \cup \mathcal{F}^1_T)$.
 For $k \in \mathcal{F}^1_T$, sensor $1$ knows the actual value of $\tau^k$ and $U^k_{\tau^k}$;  we will denote them by $t^k,u^k_{t^k}$ respectively.  
 Then,  sensor $1$'s expected cost if it decides $U^1_T =0$ is given as
\begin{align}
&\mathds{E}[J(H, U^1_T=0,\tau^1=T, (U^i_{\tau^i}, \tau^i)_{i \in \mathcal{N}}) \mid y^1_{1:T}, u^{\mathcal{P}^1}_{1:T-1}, \notag \\&~~~~~~U^1_{1:T-1}=b_{1:T-1},U^1_T=0 ] \notag \\
&= \mathds{E}[J(H,U^1_T=0,\tau^1=T, (t^k,u^k_{t^k})_{k \in \mathcal{F}^1_T}, (U^j_{\tau^j}, \tau^j)_{j \in \mathcal{G}^1_T}) \mid \notag \\&~~~~~ y^1_{1:T},u^{\mathcal{P}^1}_{1:T-1}, U^1_{1:T-1}=b_{1:T-1}] \label{eq:ap2.1}
\end{align}

We now note that given the strategies of sensors $2$ to $N$, $(U^j_{\tau^j}, \tau^j)_{j \in \mathcal{G}^1_T})$ is a deterministic function of $Y^{2:N}_{1:T}, U^1_{1:T-1}$. that is,
\begin{align}
(U^j_{\tau^j}, \tau^j)_{j \in \mathcal{G}^1_T} = \mathcal{S}(Y^{2:N}_{1:T}, U^1_{1:T-1}).
\end{align}
Therefore, the conditional expectation in \eqref{eq:ap2.1} can be written as 
\begin{align}
\sum_{h,y^{2:N}_{1:T}} \Big [& J\big(h,U^1_T=0,\tau^1=T, (t^k,u^k_{t^k})_{k \in \mathcal{F}^1_T},\mathcal{S}(y^{2:N}_{1:T},b_{1:T-1})\big) \notag \\& \times \rho^1_T(h,y^{2:N}_{1:T}) \Big]\label{eq:ap2.2}
\end{align}
Because \eqref{eq:ap2.2} is a function only  of $\rho^1_T$,$u^{\mathcal{P}^1}_{1:T-1}$  (which in turn can be computed from $\pi^1_T,u^{\mathcal{P}^1}_{1:T-1}$),  it follows that 
\begin{align}
&\mathds{E}[J(H,(U^i_{\tau^i}, \tau^i)_{i \in \mathcal{N}})| y^1_{1:T}, u^{\mathcal{P}^1}_{1:T-1},U^1_{1:T-1}=b_{1:T-1},U^1_T=0 ] \notag \\ 
& = \mathds{E}[ J(H,(U^i_{\tau^i}, \tau^i)_{i \in \mathcal{N}})|\Pi^1_T=\pi^1_T, u^{\mathcal{P}^1}_{1:T-1} ,U^1_{1:T-1}=b_{1:T-1},\notag \\&~~~~~ U^1_T=0] \label{eq:dpap1}
\end{align}
Note that the expression in \eqref{eq:dpap1} is the first term of the minimization in the definition of $V^1_T(\pi^1_T, u^{\mathcal{P}^1}_{1:T-1})$ (see \eqref{eq:dpeq1}).
Similar result holds for $U^1_T=1$ with the sensor $1$'s expected cost of deciding $U^1_T=1$ being the second term of  the minimization in the definition of $V^1_T(\pi^1_T, u^{\mathcal{P}^1}_{1:T-1})$. Therefore, for any given realization $\pi^1_T, u^{\mathcal{P}^1}_{1:T-1}$ of $\Pi^1_T, U^{\mathcal{P}^1}_{1:T-1}$, the optimal decision for sensor $1$ at time $T$ is is $0$ ($1$) if the first (second) term is the minimum in the definition of $V^1_T(\pi^1_T, u^{\mathcal{P}^1}_{1:T-1})$. Further, sensor $1$'s optimal expected cost if it hasn't stopped before time $T$ is $V^1_T(\pi^1_T, u^{\mathcal{P}^1}_{1:T-1})$.

Now assume that the theorem holds for time $t+1,t+2,\ldots,T$ and that the optimal expected cost for sensor $1$ if it hasn't stopped before $t+1$ is $V^1_{t+1}(\Pi^1_{t+1},U^{\mathcal{P}^1}_{1:t})$. 
For time $t$, using arguments similar to those used above, we can show that for $u=0,1$, sensor $1$'s expected cost if it decides $U^1_t=u$ is
\begin{align}
&\mathds{E}[J(H,(U^i_{\tau^i}, \tau^i)_{i \in \mathcal{N}})| y^1_{1:t}, u^{\mathcal{P}^1}_{1:t-1},U^1_{1:t-1}=b_{1:t-1},U^1_t=u ] \notag \\ 
& = \mathds{E}[ J(H,(U^i_{\tau^i}, \tau^i)_{i \in \mathcal{N}})|\Pi^1_t=\pi^1_t,u^{\mathcal{P}^1}_{1:t-1} ,U^1_{1:t-1}=b_{1:t-1}, \notag \\&~~~~~U^1_t=u] \label{eq:ap2.7}
\end{align}
Note that the right hand side in \eqref{eq:ap2.7} is the first term (or the second term depending on $u=0$ or $1$) of the minimization  in the definition of $V^1_t(\pi^1_t,u^{\mathcal{P}^1}_{1:t-1})$ (see \eqref{eq:dpeq2}). 

If the sensor $1$ does not stop at time $t$ (that is, it decides $U^1_t =b$), then by induction hypothesis, its optimal expected cost is 
\begin{align}
\mathds{E}[V^1_{t+1}(\Pi^1_{t+1},U^{\mathcal{P}^1}_{1:t})|y^1_{1:t}, u^{\mathcal{P}^1}_{1:t-1},U^1_{1:t-1}=b_{1:t-1},U^1_t=b ] 
\end{align}
which can be written because of Lemma \ref{lemma:update} as 
\begin{align}
&\mathds{E}[V_{t+1}(\eta_t(\pi^1_t,Y^1_{t+1},U^{\mathcal{P}^1}_t), u^{\mathcal{P}^1}_{1:t-1}, U^{\mathcal{P}^1}_{t})|y^1_{1:t}, u^{\mathcal{P}^1}_{1:t-1},\notag \\&~~~~~U^1_{1:t-1}=b_{1:t-1},U^1_t=b ] \label{eq:ap2.5}
\end{align}
We now use the fact that $U^{\mathcal{P}^1}_t$ is a deterministic function of $Y^{2:N}_{1:t}, U^1_{1:t-1}$ (see \eqref{eq:Lt}) to write \eqref{eq:ap2.5} as
\begin{align}
&\mathds{E}\big[V_{t+1}\big(\eta_t(\pi^1_t,Y^1_{t+1},\mathcal{L}_t(Y^{2:N}_{1:t},b_{1:t-1})), u^{\mathcal{P}^1}_{1:t-1}, \notag \\&~\mathcal{L}_t(Y^{2:N}_{1:t},b_{1:t-1})\big) \mid y^1_{1:t}, u^{\mathcal{P}^1}_{1:t-1},U^1_{1:t-1}=b_{1:t-1},U^1_t=b \big] \notag \\
&= \sum_{h,y^{2:N}_{1:t},y^1_{t+1}} V_{t+1}\Big(\eta_t(\pi^1_t,y^1_{t+1}, \mathcal{L}_t(y^{2:N}_{1:t},b_{1:t-1})),u^{\mathcal{P}^1}_{1:t-1},\notag \\&\mathcal{L}_t(y^{2:N}_{1:t},b_{1:t-1}) \Big)\mathds{P}(y^1_{t+1},h, y^{2:N}_{1:t}| y^1_{1:t}, u^{\mathcal{P}^1}_{1:t-1},U^1_{1:t-1}=b_{1:t-1}) \notag \\
&= \sum_{h,y^{2:N}_{1:t},y^1_{t+1}} V_{t+1}\Big(\eta_t(\pi^1_t,y^1_{t+1}, \mathcal{L}_t(y^{2:N}_{1:t},b_{1:t-1})),u^{\mathcal{P}^1}_{1:t-1},\notag \\&~~~~\mathcal{L}_t(y^{2:N}_{1:t},b_{1:t-1}) \Big)\mathds{P}(y^1_{t+1}|h)\rho^1_t(h,y^{2:N}_{1:t}) \label{eq:ap2.4}
\end{align}
Because \eqref{eq:ap2.4} is a function only  of $\pi^1_t, \rho^1_t$,$u^{\mathcal{P}^1}_{1:t-1}$  (which in turn is a function only of $\pi^1_t,u^{\mathcal{P}^1}_{1:t-1}$),  it follows that sensor $1$'s expected cost of deciding $U^1_t=b$ is
\begin{align}
&\mathds{E}[V_{t+1}(\Pi^1_{t+1},U^{\mathcal{P}^1}_{1:t})|y^1_{1:t}, u^{\mathcal{P}^1}_{1:t-1},U^1_{1:t-1}=b_{1:t-1},U^1_t=b ]  \notag \\
&= \mathds{E}[V_{t+1}(\Pi^1_{t+1},U^{\mathcal{P}^1}_{1:t})|\Pi^1_t=\pi^1_t,u^{\mathcal{P}^1}_{1:t-1} ,U^1_{1:t-1}=b_{1:T-1}, \notag \\&~~~~~U^1_t=b] \label{eq:ap2.6}
\end{align}
Note that the right hand side in \eqref{eq:ap2.6} is the third term in the minimization in definition of $V^1_t(\pi^1_t, u^{\mathcal{P}^1}_{1:t-1})$ (see \eqref{eq:dpeq2}).

Therefore, for any given realization $\pi^1_t, u^{\mathcal{P}^1}_{1:t-1}$ of $\Pi^1_t, U^{\mathcal{P}^1}_{1:t-1}$, the optimal decision for sensor $1$ at time $t$ is is $0$ (or $1$ or $b$) if the first (or second or third ) term is the minimum in the definition of $V^1_t(\pi^1_t,u^{\mathcal{P}^1}_{1:t-1})$. Further, sensor $1$'s optimal expected cost at this time is $V^1_t(\pi^1_t,u^{\mathcal{P}^1}_{1:t-1})$. This completes the induction argument.
\begin{figure*}[!t]
\begin{align}
&\mathds{E}\Big[ \mathds{E}[V^1_{t+1}(\Pi^1_{t+1},u^{\mathcal{P}^1}_{1:t-1},U^{\mathcal{P}^1}_{t}) \mid y^1_{1:t}, u^{\mathcal{P}^1}_{1:t-1}, U^{\mathcal{P}^1}_t,  U^1_{1:t-1} = b_{1:t-1},U^1_t=b] \Big|y^1_{1:t}, u^{\mathcal{P}^1}_{1:t-1},  U^1_{1:t-1} = b_{1:t-1},U^1_t=b\Big]  \label{eq:ap3.3}
\end{align} 
\line(1,0){515}
\end{figure*}
\begin{figure*}[!t]
 \begin{align}
& \sum_{y^1_{t+1}} \Big[\mathds{P}(y^1_{t+1}|H=0)\sigma^1_t(0) + \mathds{P}(y^1_{t+1}|H=1)\sigma^1_t(1)\Big] \times \notag \\&\inf_s \{\alpha^s(u^{\mathcal{P}^i}_{1:t}) \frac{\mathds{P}(y^1_{t+1}|H=0)\sigma^1_t(0)}{\mathds{P}(y^1_{t+1}|H=0)\sigma^1_t(0) + \mathds{P}(y^1_{t+1}|H=1)\sigma^1_t(1)} + \beta^s(u^{\mathcal{P}^i}_{1:t}) \label{eq:ap3.10_a}
\end{align}
 \line(1,0){515}
\end{figure*}
\begin{figure*}[!t]
\begin{align}
 & \sum_{y^1_{t+1}}\inf_s \Big\{\alpha^s(u^{\mathcal{P}^i}_{1:t}) \Big(\mathds{P}(y^1_{t+1}|H=0)\sigma^1_t(0)\Big)+\beta^s(u^{\mathcal{P}^i}_{1:t}) \Big(\mathds{P}(y^1_{t+1}|H=0)\sigma^1_t(0) + \mathds{P}(y^1_{t+1}|H=1)\sigma^1_t(1) \Big)\Big\} \label{eq:ap3.10}
 \end{align}
 \line(1,0){515}
 \end{figure*}
\section{Proof of Lemma \ref{lemma:lemma_2}}\label{sec:lemma_2}
We prove the lemma for $i=1$. We start at time $T$. Consider the first term in minimization in \eqref{eq:dpeq1}.
\begin{align}
\mathds{E}[ J(H,\{U^i_{\tau^i}, \tau^i\}_{i \in \mathcal{N}})|\pi^1_T,  u^{\mathcal{P}^1}_{1:T-1}, U^1_{1:T-1} = b_{1:T-1}, U^1_T=0] \label{eq:ap3.4}
\end{align}
Recall that  $\mathcal{F}^1_T$ is the set of sensors in $\mathcal{P}^1$ that have stopped before time $T$, that is, $\mathcal{F}^1_T = \mathcal{P}^1\setminus\mathcal{A}_T$ and  $\mathcal{G}^1_T = \mathcal{N} \setminus(\{ 1\} \cup \mathcal{F}^1_T)$.
 For $k \in \mathcal{F}^1_T$, sensor $1$ knows the actual value of $\tau^k$ and $U^k_{\tau^k}$;  we denote them by $t^k,u^k_{t^k}$ respectively. Therefore, \eqref{eq:ap3.4} can be written as
 \begin{align}
&\mathds{E}[J(H,U^1_T=0,\tau^1=T, (t^k,u^k_{t^k})_{k \in \mathcal{F}^1_T}, (U^j_{\tau^j}, \tau^j)_{j \in \mathcal{G}^1_T})) \mid \pi^1_T, \notag \\&~~~~~ u^{\mathcal{P}^1}_{1:T-1}, U^1_{1:T-1} = b_{1:T-1}, U^1_T=0] \label{eq:ap3.6}
  \end{align}
 We now note that given the strategies of sensors $2$ to $N$, $(U^j_{\tau^j}, \tau^j)_{j \in \mathcal{G}^1_T})$ is a deterministic function of $Y^{2:N}_{1:T}, U^1_{1:T-1}$. that is,
\begin{align}
(U^j_{\tau^j}, \tau^j)_{j \in \mathcal{G}^1_T} = \mathcal{S}(Y^{2:N}_{1:T}, U^1_{1:T-1}).
\end{align}
Therefore, the conditional expectation in \eqref{eq:ap3.6} can be written as 
\begin{align}
\sum_{h,y^{2:N}_{1:T}} \Big[&J\big(h,U^1_T=0,\tau^1=T, (t^k,u^k_{t^k})_{k \in \mathcal{F}^1_T},\mathcal{S}(y^{2:N}_{1:T},b_{1:T-1})\big) \notag \\&\times \rho^1_T(h,y^{2:N}_{1:T})\Big] \label{eq:ap3.5}
\end{align}
 
For any given sequence of past messages $u^{\mathcal{P}^i}_{1:T-1}$ received by sensor $1$, \eqref{eq:ap3.5} implies that the first  term in the minimization  in  definition of $V^1_T$ is  a linear function of $\rho^1_T$. Further, from \eqref{eq:ap1.6} and \eqref{eq:ap1.7} $\rho^1_T$ is an affine function of  $\pi^1_T$ and therefore the first term in minimization in $V^1_T$ is affine function of $\pi^1_T$. Similarly, the second term in the minimization  in  definition of $V^1_T$ is also an affine function of $\pi^1_T$. Thus, for a given sequence of past messages, $V^1_T(\cdot,u^{\mathcal{P}^1}_{1:T-1})$ is the minimum of two affine functions of $\pi^1_t$. This establishes the result of lemma \ref{lemma:lemma_2} for time $T$ and  also implies that  $V^1_T(\cdot,u^{\mathcal{P}^1}_{1:T-1})$is concave in $\pi^1_T$. 

Now assume that for fixed sequence of past messages $u^{\mathcal{P}^1}_{1:t}$, $V^1_{t+1}(\cdot,u^{\mathcal{P}^1}_{1:t})$ is a concave function of $\pi^1_{t+1}$. Therefore,$V^1_{t+1}(\cdot, u^{\mathcal{P}^1}_{1:t})$ can be written as infimum of affine functions of $\pi^1_{t+1}$, that is,
\begin{align}
V^1_{t+1}(\pi,u^{\mathcal{P}^1}_{1:t}) = \inf_{s}\Big\{ \alpha^{s}(u^{\mathcal{P}^1}_{1:t})\pi + \beta^s(u^{\mathcal{P}^1}_{1:t}) \Big\} \label{eq:ind_hyp}
\end{align}
where $ \alpha^{s}(u^{\mathcal{P}^1}_{1:t}),  \beta^{s}(u^{\mathcal{P}^1}_{1:t})$ are real numbers.

 We will now prove the result of the lemma for time $t$ and also establish that $V^1_t(\pi^1_t,u^{\mathcal{P}^1}_{1:t-1})$ is concave in the first argument. The first two terms in the minimization in \eqref{eq:dpeq2} are affine in $\pi^1_t$ due to arguments similar to those used for time $T$. We will prove that the third term is concave. 
Recall from \eqref{eq:ap2.6} in Appendix \ref{sec:info_states_proof} that the third term in the minimization in \eqref{eq:dpeq2} is sensor $1$'s expected cost of deciding $U^1_t=b$. 
 We can write this term as
\begin{align}
&\mathds{E}[V^1_{t+1}(\Pi^1_{t+1}, U^{\mathcal{P}^1}_{1:t})|\pi^1_t, u^{\mathcal{P}^1}_{1:t-1},  U^1_{1:t-1} = b_{1:t-1}U^1_t=b] \notag \\
&=\mathds{E}[V_{t+1}(\Pi^1_{t+1},U^{\mathcal{P}^1}_{1:t})|y^1_{1:t}, u^{\mathcal{P}^1}_{1:t-1},U^1_{1:t-1}=b_{1:t-1},U^1_t=b ].
 \label{eq:ap3.2}
\end{align} 
 Further, by smoothing property of conditional expectation, we can write \eqref{eq:ap3.2} as in \eqref{eq:ap3.3}.

 We now proceed in two steps.
 
 \textbf{Step 1:} We first consider the inner expectation in \eqref{eq:ap3.3}. Consider a realization $u^{\mathcal{P}^1}_t$. Recall that $\sigma^1_t$ is the sensor $1$'s posterior belief on $H$ and $Y^{2:N}_{1:t}$ conditioned on $y^1_{1:t}, u^{\mathcal{P}^1}_{1:t}$ and $U^1_{1:t}=b_{1:t}$--- which are precisely the terms on the right side of conditioning in the inner expectation in \eqref{eq:ap3.3}. Also, recall from \eqref{eq:ap1.9} that $\Pi^1_{t+1}$ can be obtained from $\sigma^1_t$ and $Y^1_{t+1}$; we will denote this mapping as $\Pi^1_{t+1} = \nu_t(\sigma^1_t,Y^1_{t+1})$.
 Therefore, the inner expectation in \eqref{eq:ap3.3} can be written as 
 \begin{align}
  &\mathds{E}[V^1_{t+1}(\nu_t(\sigma^1_t,Y^1_{t+1}), u^{\mathcal{P}^1}_{1:t})|\sigma^1_t,y^1_{1:t}, u^{\mathcal{P}^1}_{1:t-1}, u^{\mathcal{P}^1}_t, \notag \\&~~~~ U^1_{1:t-1} = b_{1:t-1},U^1_t=b]\label{eq:ap3.1}
 \end{align}
 The conditional expectation in \eqref{eq:ap3.1} is now only over the random variable $Y^1_{t+1}$. Note that 
\begin{align}
&\mathds{P}(Y^1_{t+1}=y^1_{t+1}| \sigma^1_t,y^1_{1:t}, u^{\mathcal{P}^1}_{1:t},  U^1_{1:t-1} = b_{1:t-1},U^1_t=b) \notag \\
&=\sum_{h} \Big[\mathds{P}(y^1_{t+1}|H=h) \mathds{P}(H=h| \sigma^1_t,y^1_{1:t}, u^{\mathcal{P}^1}_{1:t},  U^1_{1:t} = b_{1:t})\Big]
\notag \\
&= \sum_{h} \mathds{P}(y^1_{t+1}|H=h)\sigma^1_t(h), \label{eq:ap4.1}
\end{align}


 Using the induction hypothesis \eqref{eq:ind_hyp}, the form of $\nu_t$ from \eqref{eq:ap1.9} and the conditional probability of $Y^1_{t+1}$ given in \eqref{eq:ap4.1}, the right hand side of \eqref{eq:ap3.1} can be written as in \eqref{eq:ap3.10_a}, which can be further written as \eqref{eq:ap3.10}. We denote the expression in \eqref{eq:ap3.10} by the function $W^1_t(\sigma^1_t,u^{\mathcal{P}^1}_{1:t})$. Thus, $W^1_t(\sigma^1_t,u^{\mathcal{P}^1}_{1:t})$ is the inner expectation in \eqref{eq:ap3.3} for $U^{\mathcal{P}^1}_t = u^{\mathcal{P}^1}_t$. 

 Using the fact that $\sigma^1_t(1) = 1 -\sigma^1_t(0)$,  and that infimum of affine functions is concave, it follows that \eqref{eq:ap3.10} is concave in $\sigma^1_t(0)$. Therefore, $W^1_t(\sigma^1_t,u^{\mathcal{P}^1}_{1:t})$ can be written as
 \begin{equation}
 W^1_t(\sigma^1_t,u^{\mathcal{P}^1}_{1:t}) = \inf_s \{\gamma^s(u^{\mathcal{P}^1}_{1:t})  \sigma^1_t(0) +\delta^s(u^{\mathcal{P}^1}_{1:t})\}.
 \end{equation}
 
 \textbf{Step 2:} The outer expectation in \eqref{eq:ap3.3} can now be written as
 \begin{align}
 &\mathds{E}[W^1_t(\Sigma^1_t, u^{\mathcal{P}^i}_{1:t-1},U^{\mathcal{P}^i}_{t})|y^1_{1:t}, u^{\mathcal{P}^i}_{1:t-1}, U^i_{1:t-1} = b_{1:t-1},U^i_t=b] 
 \end{align}
 Note that in the above equation $\Sigma^1_t$ appearing as an argument of $W^1_t(\cdot)$ is a random variable whose realization $\sigma^1_t$ is determined after $u^{\mathcal{P}^1}_t$ is observed. The above conditional expectation can be written as  
 \begin{align}
 & \sum_{u^{\mathcal{P}^1}_t} \mathds{P}(U^{\mathcal{P}^1}_t =u^{\mathcal{P}^1}_t|y^1_{1:t}, u^{\mathcal{P}^i}_{1:t-1}, U^i_{1:t-1} = b_{1:t-1},U^i_t=b)  \notag \\
 &\times \inf_s \{\gamma^s(u^{\mathcal{P}^1}_{1:t})  \sigma^1_t(0) +\delta^s(u^{\mathcal{P}^1}_{1:t})  \} \label{eq:ap3.11}
 \end{align}
 Recall that $\rho^1_t$ is sensor $1$'s posterior belief on $H$ and $Y^{2:N}_{1:t}$ conditioned on $y^1_{1:t}, u^{\mathcal{P}^1}_{1:t-1}$ and $U^1_{1:t-1}=b_{1:t-1}$. Therefore, the conditional probability in \eqref{eq:ap3.11} can be written as
 \begin{align}
 &\mathds{P}(U^{\mathcal{P}^1}_t =u^{\mathcal{P}^1}_t|y^1_{1:t}, u^{\mathcal{P}^i}_{1:t-1}, U^i_{1:t-1} = b_{1:t-1},U^i_t=b) \notag\\
 &=\sum_{\tilde{h},\tilde{y}^{2:N}_{1:t}} \mathds{1}_{\{u^{\mathcal{P}^1}_t =\mathcal{L}_t(\tilde{y}^{2:N}_{1:t},b_{1:t-1}) \} }\times 
\rho^1_t(\tilde{h},\tilde{y}^{2:N}_{1:t}) \label{eq:ap3.12}
 \end{align}
We now use \eqref{eq:ap1.10} to write $\sigma^1_t(0)$ in terms of $\rho^1_t$. Observe that the denominator in $\sigma^1_t(\cdot)$ is exactly the same as the expression in \eqref{eq:ap3.12}. So, after some cancellations, \eqref{eq:ap3.11} reduces to infimum of affine functions of $\rho^1_t$. Because $\rho^1_t$ is affine function of $\pi^1_t$,  this   implies that \eqref{eq:ap3.11} is a concave function of $\pi^1_t$. Thus, the third term in the minimization in the definition of $V^1_t(\cdot, u^{\mathcal{P}^i}_{1:t-1})$ is concave in $\pi^1_t$ which proves the lemma for time $t$. Further,  it follows that  $V^1_t(\cdot,u^{\mathcal{P}^i}_{1:t-1})$ itself is concave in $\pi^1_t$ which completes the induction argument.

%
%
\bibliographystyle{IEEEtran}
\bibliography{myref}

\begin{thebibliography}{10}
\providecommand{\url}[1]{#1}
\csname url@samestyle\endcsname
\providecommand{\newblock}{\relax}
\providecommand{\bibinfo}[2]{#2}
\providecommand{\BIBentrySTDinterwordspacing}{\spaceskip=0pt\relax}
\providecommand{\BIBentryALTinterwordstretchfactor}{4}
\providecommand{\BIBentryALTinterwordspacing}{\spaceskip=\fontdimen2\font plus
\BIBentryALTinterwordstretchfactor\fontdimen3\font minus
  \fontdimen4\font\relax}
\providecommand{\BIBforeignlanguage}[2]{{%
\expandafter\ifx\csname l@#1\endcsname\relax
\typeout{** WARNING: IEEEtran.bst: No hyphenation pattern has been}%
\typeout{** loaded for the language `#1'. Using the pattern for}%
\typeout{** the default language instead.}%
\else
\language=\csname l@#1\endcsname
\fi
#2}}
\providecommand{\BIBdecl}{\relax}
\BIBdecl

\bibitem{Tenney_Detection}
R.~R. Tenney and {N. R. Sandell Jr.}, ``Detection with distributed sensors,''
  \emph{IEEE Trans. Aerospace Electron. Systems}, vol. AES-17, no.~4, pp.
  501--510, July 1981.

\bibitem{Tsitsiklis_survey}
J.~N. Tsitsiklis, ``Decentralized detection,'' in \emph{Advances in Statistical
  Signal Processing}.\hskip 1em plus 0.5em minus 0.4em\relax JAI Press, 1993,
  pp. 297--344.

\bibitem{Varshney}
P.~K. Varshney, \emph{Distributed Detection and Data Fusion}.\hskip 1em plus
  0.5em minus 0.4em\relax Springer, 1997.

\bibitem{Tay_tsitsiklis}
W.~P. Tay, J.~N. Tsitsiklis, and M.~Win, ``Bayesian detection in bounded height
  tree networks,'' \emph{IEEE Transactions on Signal Processing}, vol.~57,
  no.~10, pp. 4042--4051, Oct 2009.

\bibitem{Papastavrou}
J.~D. Papastavrou and M.~Athans, ``On optimal distributed decision
  architectures in a hypothesis testing environment,'' \emph{IEEE Transactions
  on Automatic Control}, vol.~37, no.~8, pp. 1154--1169, Aug 1992.

\bibitem{Tsitsiklis_large}
J.~N. Tsitsiklis, ``Decentralized detection by a large number of sensors,''
  \emph{Mathematics of Control, Signals and Systems}, vol.~1, no.~2, pp.
  167--182, 1988.

\bibitem{Chamberland04}
J.-F. Chamberland and V.~V. Veeravalli, ``Asymptotic results for decentralized
  detection in power-constrained wireless sensor networks,'' \emph{IEEE Journal
  on Selected Areas in Communication}, vol.~22, no.~6, pp. 1007--1015, Aug.
  2004.

\bibitem{Papastavrou_2}
J.~D. Papastavrou and M.~Athans, ``Distributed detection by a large team of
  sensors in tandem,'' \emph{IEEE Transactions on Aerospace and Electronic
  Systems}, vol.~28, no.~3, pp. 639--653, Jul 1992.

\bibitem{VBP_Detection}
V.~V. Veeravalli, T.~Basar, and H.~Poor, ``Decentralized sequential detection
  with a fusion center performing the sequential test,'' \emph{IEEE Trans.
  Inform. Theory}, vol.~39, pp. 433--442, Mar. 1993.

\bibitem{Ho_signaling}
Y.-C. Ho, M.~P. Kastner, and E.~Wong, ``Teams, signaling, and information
  theory,'' \emph{IEEE Transactions on Automatic Control}, vol.~23, no.~2, pp.
  305--312, Apr 1978.

\bibitem{Spence}
A.~M. Spence, \emph{Market signaling: informational transfer in hiring and
  related screening processes}.\hskip 1em plus 0.5em minus 0.4em\relax Harvard
  University Press, 1974.

\bibitem{Wald}
A.~Wald, \emph{Sequential Analysis}.\hskip 1em plus 0.5em minus 0.4em\relax
  Wiley, New York, 1947.

\bibitem{Dec_Wald}
D.~Teneketzis and Y.~C. Ho, ``The {D}ecentralized {W}ald problem,''
  \emph{Information and Computation, 73}, pp. 23--44, 1987.

\bibitem{LaVigna_86}
A.~LaVigna, {A.M. Makowski}, and {J.S. Baras}, ``A continuous-time distributed
  version of the wald's sequential hypothesis testing problem,'' \emph{Lecture
  Notes in Control and Information Sciences}, vol.~83, pp. 533--543, 1986.

\bibitem{Nayyar_MTNS}
A.~Nayyar and D.~Teneketzis, ``Decentralized detection with signaling,'' in
  \emph{Proceeding of the Workshop on the Mathematical Theory of Networks and
  Systems (MTNS)}, 2010.

\bibitem{NayyarTeneketzis:2009}
------, ``Sequential problems in decentralized detection with communication,''
  \emph{IEEE Trans. Info. Theory}, vol.~57, no.~8, pp. 5410--5435, August 2011.

\bibitem{Ho1980}
{Y. C. Ho}, ``Team decision theory and information structures,'' in
  \emph{Proceedings of the IEEE}, vol.~68, no.~6, 1980, pp. 644--654.

\end{thebibliography}

\end{document}